%% file: mst.tex
\documentclass[11pt]{article}
\RequirePackage{algorithm,algorithmic,amssymb,epsf,epic,amsmath}

\input{latexmac}
\denseformat
\usepackage{multirow}
\usepackage{boldline}

\usepackage{pdfsync,fullpage,comment}
\usepackage{authblk}

\ifx\pdftexversion\undefined
\usepackage[colorlinks,linkcolor=black,filecolor=black,citecolor=black,urlco
lor=black,pdfstartview=FitH]{hyperref}
\else
\usepackage[colorlinks,linkcolor=blue,filecolor=blue,citecolor=blue,urlcolor
=blue,pdfstartview=FitH]{hyperref}
\fi

\newcommand{\alert}[1]{\textbf{\color{red}
[[[#1]]]}\marginpar{\textbf{\color{red}**}}\typeout{ALERT:
\the\inputlineno: #1}}

\newcommand{\namedref}[2]{\hyperref[#2]{#1~\ref*{#2}}}

\begin{document}
\def\hpi{\hat{\pi}}
\def\rt{\mathit{rt}}
\def\hd{\hat{d}}
\def\chS{\hat{\cal S}}
\def\chP{\hat{\cal P}}
\def\chU{\hat{\cal U}}
\def\exp{\mathit{exp}}
\def\Rt{\mathit{Roots}}
\def\Lab{\mathit{Label}}
\def\Id{\mathit{Id}}
\def\Lev{\mathit{Level}}
\def\URt{\mathit{URoots}}
\def\Ball{\mathit{Ball}}
\def\wmax{{w_{max}}}
\def\tO{\tilde{O}}
\def\nin{\not \in}
\def\emset{\emptyset}
\def\setmns{\setminus}
\def\etal{{et al.~}}
\def\Pairs{\mathit{Pairs}}
\def\Paths{\mathit{Paths}}
\def\pred{\mathit{pred}}
\def\Rad{\mathit{Rad}}
\def\succ{\mathit{succ}}
\def\NULL{\mathit{NULL}}
\def\exp{\mathit{exp}}
\def\Ball{\mathit{Ball}}
\def\tPi{\tilde{\Pi}}
\def\deg{\mathit{deg}}
\def\TB{\cB^{{1/3}}}
\def\Branch{\mathit{Branch}}
\def\uzero{u^{(0)}}
\def\uone{u^{(1)}}
\def\uj{u^{(j)}}
\def\vzero{v^{(0)}}
\def\vone{v^{(1)}}
\def\vj{v^{(j)}}
\def\dzero{d^{(0)}}
\def\done{d^{(1)}}
\def\dj{d^{(j)}}
\def\dG{d_G}
\def\third{{1 \over 3}}
\def\stretchexp{{\log_{4/3} 7}}
\def\tLambda{\tilde{\Lambda}}
\def\tomega{\tilde{\omega}}
\def\HKNfactor{ 2^{\tO(\sqrt{\log \max \{n,\Lambda\}})}}
\def\ourfactor{ (1/\eps)^{O(\sqrt{{\log n} \over {\log\log n}})} \cdot 2^{O(\sqrt{\log n \cdot \log\log n})} }
\def\CONGEST{CONGEST}
\def\MWOE{\mathit{MWOE}}
\def\Diam{\mathit{Diam}}
\def\Id{\mathit{Id}}
\def\chF{\cal \hat{F}}

\newcommand{\Patrascu}{P\v{a}tra\c{s}cu{~}}
\newcommand{\Lists}{{\rm Lists}}
\newcommand{\td}{{\tilde{d}}}

\title{A Simple Deterministic Distributed MST Algorithm,\\  with Near-Optimal Time and Message Complexities}
\author[1]{Michael Elkin\thanks{This research was supported by the ISF grant No. (724/15).}}

\affil[1]{Department of Computer Science, Ben-Gurion University of the Negev,
Beer-Sheva, Israel. Email: \texttt{\{elkinm\}@cs.bgu.ac.il}}

\date{}
\maketitle

\begin{abstract}
Distributed {\em minimum spanning tree} (MST) problem is one of the most central and fundamental problems in distributed graph algorithms. Garay et al. \cite{GKP98,KP98} devised an algorithm with running time $O(D + \sqrt{n} \cdot \log^* n)$, where $D$ is the hop-diameter of the input $n$-vertex $m$-edge graph, and with message complexity $O(m + n^{3/2})$. Peleg and Rubinovich \cite{PR99} showed that the running time of the algorithm of \cite{KP98} is essentially tight, and asked if one can achieve near-optimal running time {\em together with near-optimal message complexity}. 

In a recent breakthrough, Pandurangan et al. \cite{PRS16} answered this question in the affirmative, and devised a {\em randomized} algorithm with time $\tO(D+ \sqrt{n})$ and message complexity $\tO(m)$. They asked if such a simultaneous time- and message-optimality can be achieved by a {\em deterministic} algorithm.

In this paper, building upon the work of \cite{PRS16}, we answer this question in the affirmative, and devise a {\em deterministic} algorithm that computes MST in time $O((D  + \sqrt{n}) \cdot \log n)$, using $O(m \cdot \log n + n \log n \cdot \log^* n)$ messages. The polylogarithmic factors in the time and message complexities of our algorithm are significantly smaller than the respective factors in the result of \cite{PRS16}. 
Also, our algorithm and its analysis are very {\em simple} and self-contained, as opposed to rather complicated previous sublinear-time algorithms \cite{GKP98,KP98,E04b,PRS16}. 
\end{abstract}


\thispagestyle{empty}
\newpage
\setcounter{page}{1}

\section{Introduction}

\subsection{Background and New Results}

Distributed {\em minimum-weight  spanning tree} (henceforth, MST) problem is one of the most fundamental and extensively studied problems in distributed graph algorithms \cite{GHS83,CT85,G85,A87,SB95,GKP98,KP98,PR99,E04,E04b,FM04,KP08,KKP11,KKT15,PRS16,MK17}.
The seminal work of Gallager et al. \cite{GHS83} gave an algorithm with running time $O(n \log n)$ and message complexity $O(m  + n \log n)$ for the problem, where $n = |V|$ and $m = |E|$ are the number of vertices and edges of the input graph $G = (V,E)$, respectively.
The time complexity was then improved to $O(n)$ \cite{CT85,G85,A87,FM04}, while still retaining the bound of $O(m + n \log n)$ on the number of messages.

Garay et al. \cite{GKP98,KP98} devised an algorithm with running time $O(D + \sqrt{n} \cdot \log^* n)$, where $D$ is the hop-diameter (equivalently, unweighted diameter) of $G$, albeit with message complexity $O(m + n^{3/2})$. 
Peleg and Rubinovich \cite{PR99} showed a lower bound of $\tilde{\Omega}(\sqrt{n})$ for the problem,\footnote{$\tO$, $\tilde{\Omega}$ and $\tilde{\Theta}$  notations hide factors polylogarithmic in $n$.}
 even when $D = O(\log n)$. 
In the open problems section of their groundbreaking paper they raised the question of devising a nearly time- and message-optimal algorithm:
\\
\\
{\em``Another research direction is to try to reduce the {\em communication} complexity of nearly time optimal algorithm of \cite{KP98} from 
$O(|E| + n^{3/2})$ towards the lower bound of $O(|E| + n \log n)$.``}
\\
\\
  In a recent breakthrough,  Pandurangan et al. \cite{PRS16} devised a randomized algorithm with time complexity $\tO(D + \sqrt{n})$ and message complexity $\tO(m)$. In the Conclusion section of their paper they write:
\\
\\
{\em ``An intriguing open question is whether randomization is necessary to simultaneously achieve time and message optimality.``}
\\
\\
In this paper we answer this question, and devise a {\em deterministic} algorithm with running time $O((D +\sqrt{n}) \cdot  \log n)$, and message complexity $O(|E| \cdot \log n + n \log n \cdot \log^* n)$. In addition to being deterministic, our algorithm is also drastically simpler than that of \cite{PRS16}.\footnote{Though we stress that it heavily builds upon several crucial ideas from \cite{PRS16}; see more details below.} Also, the polylogarithmic factors in the time and message complexities of our algorithm are significantly smaller than the respective factors in \cite{PRS16}. (Pandurangan et al. \cite{PRS16} do not specify explicitly these factors. However, since they are using an algorithm for constructing neighborhood covers from \cite{E04b}, and the latter algorithm has running time $O(D \log^3 n)$ and message complexity $O(m \cdot \log^2 n)$, these factors are definitely incurred by the algorithm of \cite{PRS16}. Also, it is apparent from their analysis that the $\sqrt{n}$ term in their time complexity is multiplied by at least $\log^2 n$.)

We also generalize our result to the $\CONGEST(b \log n)$ model, for any positive integer parameter $b$. In this model the bandwidth of every edge is $b$ edge weights and/or vertex identities.
(See Section \ref{sec:prel} for a formal definition.)
We show that our algorithm can be implemented in $O((D  + \sqrt{{n} \over b}) \cdot \log n)$ time, using $O(|E| + n \log n \cdot \log^* n)$ messages.

The lower bound for the time required to compute MST in the $\CONGEST(b \log n)$ model is $\Omega(D + \sqrt{n \over {b \log n}})$ \cite{E04,PR99}, i.e., our upper bound is $\Theta(\log n)$-off the lower bound in the first term and $\Theta(\log^{3/2} n)$-off in the second term, for all values of $b$. (In particular, this is also the gap in the standard $\CONGEST$ model, i.e., when $b = 1$.)

The lower bound on message complexity, due to Awerbuch et al. \cite{AGPV90}, is $\Omega(|E|)$.  The lower bound of \cite{AGPV90} applies to deterministic algorithms, and also even to randomized comparison-based algorithms. It also applies to randomized not comparison-based ones, as long as they apply to the so-called {\em clean network model}. In the latter model, at the beginning of the computation every vertex $v$ knows only its own identity number. On the other hand, if a vertex knows also (at the beginning of the computation) identities of all its neighbors, then a not comparison-based randomized algorithm of King et al. \cite{KKT15} achieves message complexity of $\tO(n)$ (though their time complexity is not sublinear in $n$).

Our algorithm is deterministic, comparison-based, and applies to the clean network model. (Any one of these three properties makes the lower bound of \cite{AGPV90} applicable.) 
Hence its message complexity is $O(|E| \log n + n \log n \cdot \log^* n)$ is optimal up to a $\log n$ factor in the first term, and a $\log n \cdot \log^* n$ factor in the second.

\subsection{Technical Overview}

The sublinear-time MST algorithm of \cite{KP98,GKP98} consists of two phases. In the first phase one constructs an {\em MST forest}, i.e., a collection of vertex-disjoint subtrees of the same fixed MST, that cover all vertices of the input graph $G = (V,E)$.
These subtrees are called {\em fragments}. Moreover, each of these fragments in the algorithm of \cite{KP98} has diameter $O(\sqrt{n})$, and there are $O(\sqrt{n})$ such fragments in the forest. The computation of this MST forest requires $\tO(\sqrt{n})$ time: it is done by an ingenious variant of Boruvka's algorithm. 

At this stage there are only $O(\sqrt{n})$ MST edges missing. These are  computed by a procedure, called Pipeline-MST \cite{GKP98}.
In this procedure one uses an auxiliary BFS tree $\tau$ of the input graph $G$. All candidate  edges (i.e., crossing between different fragments) are pipelined towards the root $\rt$ of $\tau$, but the key to efficiency is that every intermediate vertex $v$ of $\tau$ filters out all candidate edges $e$ that are discovered to be heaviest in some cycle. This (second) phase of the algorithm of \cite{KP98} is responsible for its large message complexity, and it also involves heavy local computations. 

The recent nearly message-optimal algorithm of \cite{PRS16} also consists of two phases, where the first phase is the same as in the algorithm of \cite{KP98}. However, on the second phase, the algorithm of \cite{PRS16} employs a different strategy than that of \cite{KP98}.
Rather than using a communication-heavy Pipeline-MST procedure, they continue merging fragments via a Boruvka-type algorithm. 

The problem with merging large-diameter fragments via Boruvka's algorithm is that  a naive implementation of this merging requires time proportional to the diameter of these fragments. When $D = O(\sqrt{n})$, Pandurangan et al. \cite{PRS16} overcome this problem by maintaining two MST forests at all times: one is the {\em base forest} $\cF$ (and its fragments are referred to as {\em base fragments}), which was computed at the first phase of the algorithm. Recall that $\cF$ consists of $O(\sqrt{n})$ fragments of size $O(\sqrt{n})$ each.
(For a pair of parameters $\alpha,\beta$, an {\em $(\alpha,\beta)$-MST forest} is an MST forest with at most $\alpha$ fragments, each of diameter at most $\beta$. The base forest is an $(O(\sqrt{n}),O(\sqrt{n}))$-MST forest.) 

The second MST forest $\chF$ that the algorithm of \cite{PRS16} maintains is obtained by merging some of the base fragments into fragments of $\chF$ via Boruvka's algorithm. To compute the  {\em minimum weight outgoing edge} (henceforth, MWOE) of a fragment $\hF \in \chF$, the algorithm computes in each base fragment $F \in \cF$ a minimum-weight edge $e_F$ crossing between $V(F)$ and $V \setminus V(\hF)$, where $F \subseteq \hF$, and $\hF \in \chF$. Then the algorithm upcasts these edges $e_F$ to the root $\rt$ of the auxiliary BFS tree $\tau$. The root $\rt$ uses this information to compute the MWOE $e_{\hF}$ of  every fragment $\hF \in \chF$, and then to compute a new MST forest $\chF'$. The fragments of the latter forest are obtained by merging some of the fragments of $\chF$, via Boruvka's algorithm. 

When $D \le \sqrt{n}$, the procedure described above is both time- and message-efficient. However, generally, its message complexity is $\tilde{\Theta}(D \sqrt{n} + n)$, and this is super-linear for $D = \omega(\sqrt{n})$. To resolve this issue, \cite{PRS16} employ hierarchies of sparse neighborhood covers \cite{ABCP93,C93,E04b}, and use them build what they call ``communication-efficient fragments and paths'' within large-diameter fragments. This results in a sophisticated and complicated algorithm, with an elaborate analysis, which incurs quite a few polylogarithmic factors in both time and message complexities, and requires storing certain non-trivial local data structures in every vertex.
Moreover, since there are currently no known {\em deterministic} distributed time- and message-efficient algorithms for constructing neighborhood covers, the algorithm of \cite{PRS16} resorts to using a {\em randomized} algorithm of \cite{E04b}. As a result, the solution of \cite{PRS16} becomes randomized as well. 

In this paper we propose a different, and a much simpler solution, for the situation when $D \ge \sqrt{n}$. Instead of constructing an $(O(\sqrt{n}),O(\sqrt{n}))$-MST forest $\cF$ as a base forest, we construct an $(O(n/D),O(D))$-MST forest. By slightly generalizing and refining the analysis of \cite{KP98,L16,PRS16}, we show that this can be done in $O(D \cdot \log^* n)$ time, and with $O(|E| \cdot \log D + n \cdot \log D \cdot \log^* n)$ messages. By doing so we spend more time on the first phase than the algorithms of \cite{KP98,PRS16}; this is however  still well within our desired time bounds.

Then we use the algorithm of \cite{PRS16} on top of this base forest, as opposed to using it on top of an $(O(\sqrt{n}),O(\sqrt{n}))$-MST forest. (Recall that, as was argued above, the latter would have not been message-efficient.) Now computing a minimum-weight edge $e_F$ crossing between the vertex set $V(F)$ of a base fragment $F$ and $V \setmns V(\hF)$, where $\hF \in \chF$ is the fragment that contains $F$, can be done in $O(D + n/D) = O(D)$ time. Even more importantly, upcasting all these edges $e_F$ to the root $\rt$ of the auxiliary tree $\tau$  requires now just $O(D \cdot n/D) = O(n)$ messages. As a result, the entire message complexity of our algorithm is near-linear. 

As opposed to previous solutions, our entire algorithm and its analysis are ultimately very simple. In fact, we essentially provide all the details (including those which originate from previous work) in this extended abstract. 

\subsection{Related Work}

Singh and Bernstein \cite{SB95} devised an MST algorithm with near-optimal message complexity, and with running time $O((\Delta + \Diam(MST)) \cdot \log n)$, where $\Delta$ is the maximum degree of the input graph $G = (V,E,\omega)$, \footnote{$\omega:E \rightarrow \mathtt{R}^+$ is a weight function on edges of $G$.} and $\Diam(MST)$ is the hop-diameter of the computed MST of $G$. The latter parameter is always greater or equal to $D = \Diam(G)$, but for many instances it is smaller than $n$.

The current author \cite{E04b} devised an MST algorithm with running time $\tO(\mu(G,\omega) + \sqrt{n})$, where $\mu(G,\omega)$ is a parameter which is never greater than $D$, and for many instances it is much smaller than $D$. There is also  a lower bound of $\Omega(\mu(G,\omega)) + \tilde{\Omega}(\sqrt{n})$ for the MST computation on an input graph $(G,\omega)$ \cite{E04b}. 
Albeit, the algorithm of \cite{E04b} does not detect termination (unless it is given an estimate of $\mu(G,\omega)$ as a part of the input).

Khan and Pandurangan \cite{KP08} devised an $O(\log n)$-approximate MST algorithm with running time $\tO(D + L(G,\omega))$, where $L(G,\omega)$ is yet another parameter, called {\em local shortest path diameter}. It may be smaller or larger than $D$.

Lower bounds on the time required to compute an approximate MST were shown in \cite{E04,SHKKNPPW12,EKNP14}. In particular, \cite{EKNP14} showed such lower bounds even when quantum distributed communication is allowed.
Lower bounds for MST on graphs with constant hop-diameter $D$ were shown in \cite{LPP06,E04}.
MST on graphs with $D=1$ (the {\em Congested Clique} model) was studied in \cite{LPPP05,HPPSS15,GP16}. In particular, \cite{HPPSS15} devised a message-optimal and time-efficient MST algorithm for this model.

Mahreghi and King \cite{MK17} devised randomized, not comparison-based MST algorithm with running time $\tO(\Diam(MST))$, and with $\tO(n)$ messages. (This algorithm assumes that at the beginning of the computation, every vertex knows the identities of all its neighbors, i.e., the so-called $KT_1$ model.)

Efficient MST algorithms for planar graphs,  and more generally, graphs of bounded genus, were given in \cite{GH16,HIZ16}.

\section{Preliminaries}
\label{sec:prel}

We consider the synchronous $\CONGEST$ model of distributed communication. Every vertex $v$ of an input graph $G = (V,E)$ hosts a processor, and these processors communicate with one another via $O(\log n)$-size messages in synchronous rounds.
All edge weights are assumed to be at most polynomial in $n$, or alternatively, the message size can be restricted to $O(1)$ edge weights 
or/and identity numbers.
In a more general $\CONGEST(b \log n)$ model, for a parameter $b \ge 1$, 
on every round every vertex is allowed to send messages of size $O(b \log n)$ bits, or alternatively, $O(b)$ edge weights and/or vertex identities via every edge incident on it.

At the beginning of the communication every vertex $v$ knows its own unique identity number, denoted $\Id(v)$. 
The {\em running time} of an algorithm in this model is the worst-case number of rounds that it runs. The {\em message complexity} of an algorithm  is the worst-case overall number of messages sent throughout an execution of the algorithm.
At the end of an execution, every vertex $v$ is required to know which among the edges incident on it belong to the MST.

We assume that the MST is unique. This assumption is without loss of generality, see, e.g., \cite{Pel00:ln}, Ch. 5. A connected subtree of the unique MST is called an {\em MST fragment}, or simply a {\em fragment}.

We say that a collection $\{F_1,F_2,\ldots,F_h\}$, for some positive integer $h$, is an {\em MST forest}, if for each $i \in [h]$,  $F_i$ is an MST fragment, these fragments are vertex-disjoint, and $\bigcup_{i =1}^h V(F_i) = V$.
For a pair of positive  parameters $\alpha$ and $\beta$, we say that an MST forest $\cF$ 
is an {\em $(\alpha,\beta)$-MST-forest}, if it contains at most $\alpha$ fragments, each with strong diameter at most $\beta$. 
({\em Strong diameter} of a subgraph $F$ is the maximum distance in $F$ between a pair of vertices $u,v \in V(F)$.)
A {\em diameter} of an MST forest $\cF$ is the maximum diameter of one of its fragments.

We say that an MST forest $\cF'$ {\em coarsens} MST forest $\cF$, if for every fragment $F \in \cF$, there exists a fragment $F' \in \cF'$ that contains it, i.e., $V(F) \subseteq V(F')$ (and, as a result, also $E(F) \subseteq E(F')$, because $F$ and $F'$ are subtrees of the same spanning tree). 

Boruvka's algorithm starts from a collection of MST fragments. On each phase it computes the  MWOE of every fragment, and computes the fragments' graph, whose vertices are the fragments, and edges are the MWOEs. It then merges each connected component of the fragments' graph into a greater fragment, and obtains an MST forest with fewer fragments. In fact, the number of fragments decreases at least by a factor of 2, and so the number of phases is $O(\log n)$. See \cite{Pel00:ln}, Ch. 5, for further details.

For a rooted tree $T$ and a non-root vertex $v$ in $T$, we denote by $\pi_T(v)$ the {\em parent} of $v$ in $T$.
For a vertex $v$, we denote by $\Id(v)$ the identity of the vertex $v$. For each fragment $F$, there is a designated root vertex $\rt_F$, and the identity $\Id(F)$ of $F$ is set  to be the identity $\Id(\rt)$ of the root $\rt$.

\section{The Algorithm and its Analysis}
\label{sec:alg}

Based on \cite{GKP98,KP98} (see also \cite{PRS16}, Algorithm 1, called Controlled-GHS, and Lemma 1, and Lenzen's lecture notes \cite{L16}, the chapter about MST, Lemmas 6.15-6.17), we show in Section \ref{sec:forest}
 that for any positive parameter $k$, an $(n/k,O(k))$-MST forest  $\cF = \cF_0$ can be computed in $O(k \cdot \log^*n)$ time, and using $O(|E| \log k + n \log k \cdot \log^* n)$  messages. \footnote{In fact, Lemma 6.17 of \cite{L16} applies this only for $k \le \sqrt{n}$, but  inspecting its proof reveals that it holds for larger values of $k$ as well. The message complexity of this procedure is not analyzed in \cite{KP98,L16}, while its analysis in \cite{PRS16} provides a slightly weaker bound. For the sake of completeness, we provide a self-contained proof of this result in Section \ref{sec:forest}.}
 We refer to $\cF_0$ as the {\em base} MST forest, and call its fragments {\em base fragments}.

The algorithm starts with constructing an auxiliary BFS tree $\tau$ for the entire graph $G$ rooted at a root vertex $\rt$. This step requires $O(D)$ time and $O(|E|)$ messages.

Every base fragment $F$ has its designated root vertex $r_F$. We need every vertex $v$ of $\tau$ to be able to route messages from the root $\rt$ of $\tau$ to each of the roots $r_F$ of base fragments $F \in \cF$, which belong to the subtree $\tau_v$ of $\tau$ rooted at $v$.
 For this end, we compute intervals $I_v$ for each vertex $v \in V(\tau)$, such that for every pair $u,v$ of vertices in $V$, 
their intervals are either disjoint (if they belong to different branches of $\tau$), or nested if the vertex with a larger interval is an ancestor in $\tau$ of the vertex with a smaller interval. Given these intervals, when a vertex $v$ needs to route a message to a root $r_F$ of a base fragment $F$ which belongs to $V(\tau_v)$, it finds a child $u$ of $v$ whose interval $I(u)$ contains $I(r_F)$, and sends the message to this child.

To compute the intervals, we first conduct a convergecast in $\tau$. As a result of this convergecast, every vertex $v$ knows the size $|V(\tau_v)|$ of its subtree. Then the root $\rt$ of $\tau$ assigns itself the interval $I(\rt)= [1,n]$, $n = |V(\tau)| = |V|$, and assigns its children $u_1,\ldots,u_d$, for $d = \deg(\rt)$, disjoint intervals $I(u_1),\ldots,I(u_d) \subseteq I(\rt)$, with $|I(u_i)| = |V(\tau_{u_i})|$, for every $i \in [d]$.
(This is possible because $\sum_{i=1}^d |V(\tau_{u_i})| = n-1 = |I(\rt)| - 1$. Observe also that $\rt$ can learn $|V(\tau_{u_i})|$, for all $i \in [d]$, within one round.) Next, each of the children $u_i$ assigns (in parallel) disjoint intervals to their children, etc.
Finally, at the end of this process, we conduct a pipelined convergecast during which the root $\rt$ learns the $|\cF|$ intervals of all the base fragments.

    The entire process of computing the intervals requires $O(D)$ time and $O(n)$ messages, while the final pipelined convergecast requires $O(D + |\cF|) = O(D + n/k)$ time, and $O(D \cdot n/k)$ messages.

Consider first the case $D \le \sqrt{n}$. We set $k =  \sqrt{n}$.
Suppose we have already conducted $j$ phases of the Boruvka's algorithm, starting from $\cF_0$, and obtained a coarsening forest $\cF_j$, for some $j = 0,1,2,\ldots$, of $\cF$. 
We now show how to implement the next phase of Boruvka's algorithm, and to construct a coarsening MST forest $\cF_{j+1}$ of $\cF_j$ (and, consequently, of $\cF$ too).

We assume that every vertex $v$ knows the identities of both the base fragment $F_v$ and the fragment $\hF_v$ of $\cF_j$ that it belongs to. 
Also, for every neighbor $u$ of $v$, we assume that $v$ knows the identities of $F_u$ and $\hF_u$.
We also assume that the root $\rt$ knows the identities of all base fragments,  and at the beginning of phase $j$, $j = 0,1,\ldots$, it knows the identities of all fragments of $\cF_j$, and for each base fragment $F \in \cF$, the root knows the identity of the fragment $\hF \in \cF_j$ that coarsens it.  This is argued by induction on $j$.

To guarantee that the induction base $j = 0$ holds, after the base MST forest is constructed, every vertex $v$ updates its neighbors with the identity of $F_v$. This requires $O(1)$ time and $O(|E|)$ messages. Also, an upcast of $|\cF_0| \le n/k$ identities of base fragments is conducted over the BFS tree $\tau$ at this stage. This step requires $O(D + n/k)$ time, and $O(D \cdot n/k)$ messages.

In every base fragment $F \in \cF_0$ we compute (in parallel in all base fragments) the edge $e = (u,v)$ of minimum weight that crosses between $u \in V(F)$ and $v \in V \setminus V(\hF)$, where $\hF \in \cF_j$ is the fragment that coarsens the base fragment $F$. This computation requires $O(k) = O(\sqrt{n})$ time, and $O(n)$ messages.

Once this is done, we upcast all these $O(n/k) = O(\sqrt{n})$ pieces of information over the auxiliary 
BFS tree $\tau$ to the root vertex $\rt$ of $\tau$. 
This is done via a pipelined convergecast procedure, in which every intermediate vertex $u$ of $\tau$ forwards to his parent $\pi_\tau(u)$ in $\tau$ only the lightest edge for each fragment $\hF \in \cF_j$, among edges that were initially stored at one of the vertices $z$ of the subtree $\tau_u$ of $\tau$, rooted at $u$. 
This step requires $O(D + |\cF_j|)$  time,
and $O(D \cdot |\cF_j|)$ messages.  (See \cite{Pel00:ln}, ch. 3.)

The root $\rt$ locally computes the MWOE $e_{\hF}$ for every fragment $\hF \in \cF_j$. It then locally  computes the fragments' graph whose vertices are fragments of $\cF_j$, and edges are the MWOEs, and computes the MST forest $\cF_{j+1}$.
Specifically, for every base fragment $F \in \cF$, the root knew the identity of a fragment $\hF \in \cF_j$ that coarsens it. 
As a result of the computation that $\rt$ conducts, it now knows the identity of a fragment $\hF' \in \cF_{j+1}$ that coarsens $\hF$.  (Consequently, $\hF'$ also coarsens $F$.)
The root $\rt$ then sends $|\cF|$ messages over $\tau$, each message is of the form $(F,\hF')$, where  $F \in \cF$, $\hF' \in \cF_{j+1}$,
$\hF'$ coarsens $F$. 
Each such a message  $(F,\hF')$ has the destination interval $I(\rt_F)$ attached to it, and it is routed along the unique $\rt-\rt_F$ path in $\tau$. 
The  root $\rt_F$ of the base fragment $F$ receives this message, and
  writes down to itself that it belongs to $\hF'$.
This (pipelined) downcast requires $O(D + |\cF|)$  time, and $O(D \cdot |\cF|) = O(D \cdot n/k)$ messages.
(This is because every one of the $|\cF|$ messages is routed to its destination along a path with at most $D$ edges.)

Next, every root vertex $r_F$ of a base fragment $F \in \cF$ broadcasts the identity $\Id(\hF')$ of their new $(j+1)$st level fragment $\hF' \in \cF_{j+1}$ to all vertices of $F$.
This requires $O(k)$ time and $O(n)$ messages. Finally, every vertex $v$ updates its neighbors in $G$ with its new $(j+1)$st level's fragment identity. This requires $O(1)$ time, and $O(|E|)$ messages.
 This completes the description of a single phase of Boruvka's algorithm.

To analyze the running time and message complexity, observe that for every $j = 0,1,2,\ldots$, we have $|\cF_{j+1}| \le \half \cdot |\cF_j|$, and so the number of phases $\ell$  is $O(\log n)$ phases. Hence the overall time is 
\begin{equation}
 O(D + n/k) + O(k \cdot \log^* n) + O((D + k + |\cF|) \cdot \log n) ~=~ O((D + k + n/k) \cdot \log n) ~=~ O(\sqrt{n}\cdot \log n).
\end{equation}
Similarly, the message complexity is $O(|E| \log n + n \log n \cdot \log^* n)$ for constructing $\cF$, $O(D \cdot n/k + n)$ for computing the intervals,
  and  $O(D \cdot n/k + |E| + n)$ on each consequent phase.
As $D \le k$, 
the overall message complexity is $O(|E| \log n + n\log n \cdot \log^* n)$. 

For $D > \sqrt{n }$, we compute the $(n/k, O(k))$-MST forest $\cF = \cF_0$ with parameter $k = D$ in $O(D \cdot \log^*n)$ time, and
$O(|E| \log n + n\log n \cdot \log^* n)$ messages. From this point on, the algorithm is identical to the one that we have just described.
For every $j = 0,1,2,\ldots$, the $j$th phase of it requires $O(D + k + |\cF|) = O(D + k + n/k) = O(D)$ time, i.e., all phases altogether require $O(D \log n)$ time.

 The number of messages  is $O(|E| +  n + D \cdot |\cF|) $ on every phase,  i.e., $O((|E| + n) \cdot \log n)$ 
messages in all the $\ell$ phases.
Hence the total message complexity is $O(|E| \log n + n\log n\cdot \log^* n)$.

We summarize this result below.

\begin{theorem}
\label{thm:mst}
The deterministic algorithm that was described above computes the minimum spanning tree in the $\CONGEST$ model, in $O((D  + \sqrt{n}) \cdot \log n)$ time, using $O(|E| \log n + n \log n \cdot \log^* n)$ messages.
\end{theorem}

Next, we extend the algorithm to the $\CONGEST(b \log n)$ model, for a positive integer parameter $b$. 
We first discuss the case of small diameter, i.e., $D \le \sqrt{n \over {b }}$, and then proceed to discussing the complementary case. 

In the small-diameter regime, we set $k =  \sqrt{n \over {b }}$, i.e., $D \le k$. We construct an $(n/k,O(k))$-MST forest $\cF_0$ in $O(k \log^* n)$ time, using $O(|E| \log n + n\log n \cdot \log^* n)$ messages. 
The upcast of $|\cF_0| \le n/k$ identities of base fragments requires $O(D + {n \over {k \cdot b}})$ time and $O(D \cdot n/k)$ messages.
Now consider the $j$th phase of the algorithm, for some $j = 0,1,2,\ldots$.
Computing minimum weight crossing edges in parallel in all base fragments
$\{e = (u,v) \mid u \in V(F), v \in V \setminus V(\hF),  F \in \cF, \hF \in \cF_j\}$ requires $O(k)$ time and $O(n)$ messages.
Pipelined convergecast of $|\cF_j|$ items requires $O(D + {{|\cF_j|} \over b})$ time and $O(D \cdot |\cF_j|)$ messages.
The pipelined downcast of $|\cF| \le n/k$  messages requires $O(D + |\cF|/b) = O(D + {n \over {k b}}) = O(D + \sqrt{n/b})$ time,  and $O(D \cdot |\cF|) = O(D \cdot n/k)$ messages.
(Note that this downcast sends each message only along its own root-destination path, rather than broadcasting it to the entire graph.)
Updating neighbors with new fragments' identities requires $O(1)$ time and $O(|E|)$ messages.
The overall running time of the $\ell$ phases is 
\begin{eqnarray*}
&& O\left(D + {n \over {k \cdot b}}\right) + O(k \log n + D \log n +   |\cF| \cdot \log n)  ~=~ O\left(\left(D+ k + {n \over {k \cdot b}}\right) \cdot \log n\right) \\
&&=~
O\left(\sqrt{{n } \over b} \cdot \log n\right)~.
\end{eqnarray*}
This is also the upper bound on the total running time. The overall number of messages used in the $\ell$ phases is
$O((D \cdot n/k  + n+  |E|) \log n) = O(|E| \cdot \log n)$.
Hence the total message complexity is $O(|E| \log n + n \log n \cdot \log^* n)$.

In the large-diameter  regime, i.e., when $D > \sqrt{n \over b}$, we set $k = D$.
Constructing $\cF = \cF_0$ requires $O(D \log^* n)$ time and $O(|E| \log n + n \log n \cdot \log^* n)$ messages.
Computing minimum weight crossing edges in all base fragments in parallel requires  (on each phase) $O(D)$ time and $O(n)$ messages.
Other than that on phase $j$, for $j = 0,1,\ldots,\ell-1$, we have time $O(D + {{|\cF|} \over b})$ time and $O(D \cdot |\cF|)$ messages.
Overall, this sums up to
$$O((D + |\cF|/b) \cdot \log n) ~=~ O\left(\left(D + {n \over {D b}}\right) \cdot \log n\right) ~=~ O(D\log n)$$ time,
and $O( (D \cdot n/k + n + |E|) \cdot \log n) = O(|E| \cdot \log n)$  messages.
Hence the total running time of the entire algorithm in this case is $O(D \log n)$, and its message complexity is $O(|E|\log n + n \log n \cdot \log^* n)$.

\begin{theorem}
\label{thm:congestb}
For any $b \ge 1$, the deterministic algorithm that was described above computes the minimum spanning tree in $\CONGEST(b \log n)$ model, in $O((D  + \sqrt{n/b})\cdot \log n$ time, using $O(|E| \log n + n \log n \cdot \log^* n)$ messages.
\end{theorem}

\section{Constructing an MST Forest}
\label{sec:forest}

For the sake of completeness, we next describe the algorithm (due to \cite{GKP98,KP98,L16}) for constructing an $(n/k,O(k))$-MST forest, for an integer parameter $k \le n/10$. (The constant 10 is quite arbitrary.) Our version of the algorithm is slightly more general than that in \cite{L16}, and our bounds on its time and message complexities are slightly better than the respective bounds in \cite{PRS16}. 

The algorithm runs for $t = \lceil \log k\rceil$ phases. At the beginning of a phase $i = 0,1,2,\ldots,t-1$, the algorithm has already computed
$(n/2^{i-1},6 \cdot 2^i)$-MST forest $\cF_i$. The induction base $i=0$ holds for the MST forest of singletons.
We next describe a single phase of the algorithm, and show that the resulting collection $\cF_{i+1}$ is an $(n/2^i,6\cdot 2^{i+1})$-MST forest.

At the beginning of a phase $i$, every fragment $F \in \cF_i$ with diameter at most $2^i$ computes the edge $e_F = \MWOE(F)$. We denote the set of fragments $\cF'_i$. This step requires $O(2^i)$ time and $O(n)$ messages.
(Also, at the beginning of each phase, every vertex updates its neighbors with the identity of its fragment. This requires $O(1)$ time and $O(|E|)$ messages.)
Then, for every $e_F = (u,v)$, $u \in V(F)$, $v \in V \setminus V(F)$, a message is sent over $e_F$, and the receiver $v$ writes down $u$ as a ``foreign-fragment''  child of itself. (In the special case when $(u,v) = \MWOE(F_u) = \MWOE(F_v)$, with $u \in V(F_u)$, $v \in V(F_v)$, the endpoint belonging to a higher-identity fragment becomes the parent of the other endpoint.)

This defines a {\em candidate fragment graph} $\cG'_i = (\cF'_i,\cE_i)$, whose vertices are the fragments of $\cF'_i$, and edges are the $\MWOE$ edges of these fragments. We then compute a maximal matching (henceforth, MM) $M$ in $\cG'_i$.
(We will soon elaborate on this.) For every pair $(F,F') \in M$, the two fragments merge into a single fragment, along the $\MWOE$ edge that connects them.
Every fragment $F'' \in \cF_i \setminus \cF'_i(M)$ is necessarily connected via its $\MWOE(F'') = e''$ to either a matched fragment $F \in \cF'_i(M)$, or to a fragment $F \in \cF_i \setminus \cF'_i$ of diameter larger than $2^i$. In either case, it now merges with $F$ along the edge $e''$. Except for the procedure that computes an MM, this completes the description of the algorithm.
The MST forest $\cF_{i+1}$ consists now of the resulting merged fragments, and of those fragments of $\cF_i \setminus \cF'_i$ that did not participate in the merging process described above. (These are the fragments $F$ with diameter $\Diam(F) > 2^i$, and such that no unmatched fragment $F' \in \cF'_i$ has its $\MWOE$ $(u,v)$ with an endpoint in $F$.) 

\inline Remark: The next two lemmas and their proofs are closely related to that of Lemmas 6.15 and 6.17 in \cite{L16}.

\begin{lemma}
\label{lm:diam}
$\Diam(\cF_{i+1}) \le 6 \cdot 2^{i+1}$.
\end{lemma}
\proof
Each new fragment $\hF \in \cF_{i+1}$ can be viewed as a subtree of diameter at most 3 in the fragment graph $\cG_i = (\cF_i,\cE_i)$, whose edge set $\cE_i$ is the set of all the $\MWOE$s of fragments of $\cF_i$. Moreover, at most one of the fragments in this subtree may have diameter greater than $2^i$. (But, by induction hypothesis, its diameter is, nevertheless, at most $6 \cdot 2^i$.)
Hence the diameter of $\hF$ in $G$ is at most $6 \cdot 2^i + 3 \cdot 2^i + 3 \le 12 \cdot 2^i = 6 \cdot 2^{i+1}$.
\QED
 
\begin{lemma}
\label{lm:size}
For $i = 0,1,\ldots,t-2$, each fragment $\hF \in \cF_{i+1}$ contains at least $2^i$ vertices.
\end{lemma}
\proof
The proof is by induction on $i$. The base $i = 0$ holds, as every fragment of $\hF_1$ contains at least one vertex.

By induction hypothesis, every fragment $F \in \cF_i$ contains at least $2^{i-1}$ vertices. Consider a fragment $F \in \cF_i$ with $|F| < 2^i$. 
Then $\Diam(F) < 2^i$ too, and hence $F \in \cF'_i$. Thus $F$  merges with at least one other fragment $F'$ of $\cF_i$.
(As $|F| < 2^i \le n$, $F$ has outgoing edges, and thus has an $\MWOE$.)
Since, by induction hypothesis, $|F|,|F'| \ge 2^{i-1}$, it follows that the merged fragment has size at least $2^i$.
\QED

Hence $|\cF_i| \le n/2^{i-1}$. By substituting $i = t-1 = \lceil \log k\rceil -1$, we get $|\cF_{t-1}| = O(n/k)$, and $\Diam(F_{t-1}) = O(k)$.
By rescaling (setting $k' = c \cdot k$, for an appropriate constant $c$), we obtain the desired $(n/k,O(k))$-MST forest.

Next, we sketch the procedure that computes an MM in the forest $\cG'_i = (\cF'_i,\cE'_i)$, and analyze its time and message complexities.
Recall that, by Lemma \ref{lm:diam}, each fragment $F \in \cF'_i$ has diameter $O(2^i)$.

The first step is to simulate Cole-Vishkin's 3-vertex-coloring algorithm \cite{CV86} in $\cG'_i$. For this end, on every step every internal fragment $F \in \cF'_i$ needs to send a message with its current color to its children in $\cG'_i$.
(Initial colors are set as fragments' identities.)
 This is implemented in $O(2^i)$ time, using $O(n)$ messages, in a straightforward manner. Since there are $\log^* n$ such steps, overall this computation requires $O(2^i \cdot \log^* n)$ time and $O(n \cdot \log^* n)$ messages. 

Given a 3-vertex-coloring of $\cG'_i$, there are 3 steps. On each step $j \in \{1,2,3\}$, fragments $F'$ of color $j$ that have at least one of their children $F''$ unmatched, insert an edge $(F',F'')$ connecting them to such an unmatched child into the matching, and update their parents $F$ that they became matched. The second step involves a convergecast in the parent fragment $F$, during which the root of $F$ learns if it still has an unmatched child. Also, every internal vertex $v \in F$ learns if one of its descendents leads to an unmatched child of $F$ or not.

This entire part of the algorithm can also be implemented in $O(2^i)$ time and $O(n)$ messages, in a straightforward manner.
Hence the entire computation of MM on phase $i$ of the algorithm requires $O(2^i \cdot \log^*n)$ time, and $O(n \cdot \log^* n)$ messages.
Thus, the total running time of phase $i$ is $O(2^i \cdot \log^* n)$, and the number of messages if $O(|E| + n \cdot \log^* n)$.
Summing up over all the $\lceil \log k\rceil$ phases, we obtain running time $O(\log^* n \sum_{i=0}^{\lceil \log k\rceil} 2^i)  = 
O(k \cdot \log^* n)$, and message complexity $O(|E| \cdot \log k + n  \log k \cdot \log^* n)$.

We summarize this section in the following theorem.

\begin{theorem}
\label{thm:forest}
For an integer parameter $k \le n/10$, the deterministic algorithm described above computes
 an $(n/k,O(k))$-MST forest in time
$O(k \cdot \log^* n)$, using $O(|E| \cdot \log k + n \cdot \log k \cdot \log^* n)$ messages.
\end{theorem}

\newpage
\pagenumbering{gobble}
\bibliographystyle{alpha}
\bibliography{mst}

\end{document}

%% file: latexmac.tex
\def\denseformat{
\setlength{\textheight}{9in}
\setlength{\textwidth}{6.9in}
\setlength{\evensidemargin}{-0.2in}
\setlength{\oddsidemargin}{-0.2in}
\setlength{\headsep}{10pt}
\setlength{\topmargin}{-0.3in}
\setlength{\columnsep}{0.375in}
\setlength{\itemsep}{0pt}
}

\newtheorem{theorem}{Theorem}[section]

\newtheorem{lemma}[theorem]{Lemma}

\newtheorem{comment}[theorem]{Comment}

\def\boldhead#1:{\par\vskip 7pt\noindent{\bf #1:}\hskip 10pt}
\def\ithead#1:{\par\vskip 7pt\noindent{\it #1:}\hskip 10pt}

\def\inline#1:{\par\vskip 7pt\noindent{\bf #1:}\hskip 10pt}
\def\midinline#1:{\par\noindent{\bf #1:}\hskip 10pt}
\def\dnsinline#1:{\par\vskip -7pt\noindent{\bf #1:}\hskip 10pt}
\def\ddnsinline#1:{\newline{\bf #1:}\hskip 10pt}
\def\largeinline#1:{\par\vskip 7pt\noindent{\large\bf #1:}\hskip 10pt}
\long\def\comment #1\commentend{}
\long\def\commhide #1\commhideend{}
\long\def\commfull #1\commend{#1}
\long\def\commabs #1\commenda{}
\long\def\commtim #1\commendt{#1}
\long\def\commb #1\commbend{}
\long\def\commedit #1\commeditend{} 

\long\def\commB #1\commBend{}       

\long\def\commex #1\commexend{}     

\long\def\commsiena #1\commsienaend{}  
                                         
\long\def\commBI #1\commBIend{}  
                                         
\long\def\CProof #1\CQED{}

\def\blackslug{\hbox{\hskip 1pt \vrule width 4pt height 8pt
    depth 1.5pt \hskip 1pt}}
\def\QED{\quad\blackslug\lower 8.5pt\null\par}

\def\Proof{\par\noindent{\bf Proof:~}}

\def\proof{\Proof}

\long\def\PPP#1{\noindent{\bf Proof:}{ #1}{\quad\blackslug\lower 8.5pt\null}}

\long\def\denspar #1\densend
{#1}

\newif\ifnotesw\noteswtrue
   {\ifnotesw\marginpar[\hfill\(\top\)]{\(\top\)}\fi}%
   {\ifnotesw\marginpar[\hfill\(\bot\)]{\(\bot\)}\fi}

\newcommand{\mnote}[1]%
    {\ifnotesw\marginpar%
        [{\scriptsize\it\begin{minipage}[t]{\marginparwidth}
        \raggedleft#1%
                        \end{minipage}}]%
        {\scriptsize\it\begin{minipage}[t]{\marginparwidth}
        \raggedright#1%
                        \end{minipage}}%
    \fi}

\def\cB{{\cal B}}

\def\cE{{\cal E}}
\def\cF{{\cal F}}
\def\cG{{\cal G}}

\def\hF{{\hat F}}

\def\tO{{\tilde O}}

\def\MathF{\hbox{\rm I\kern-2pt F}}
\def\MathP{\hbox{\rm I\kern-2pt P}}
\def\MathR{\hbox{\rm I\kern-2pt R}}
\def\MathZ{\hbox{\sf Z\kern-4pt Z}}
\def\MathN{\hbox{\rm I\kern-2pt I\kern-3.1pt N}}
\def\MathC{\hbox{\rm \kern0.7pt\raise0.8pt\hbox{\footnotesize I}
\kern-4.2pt C}}
\def\MathQ{\hbox{\rm I\kern-6pt Q}}

\newsavebox{\ttop}\newsavebox{\bbot}

\def\eps{\epsilon}

\def\setmns{\setminus}

\newcommand{\half}{{1\over2}}

\def\nin{{~\not \in~}}
\def\emset{\emptyset}

\def\etal{\emph{et~al.}}
